# Systematic Mapping Protocol

UX Design Role in Software Development Process

Final version: 20/02/2024

Emilio Gustavo Ormeño, Fernando Pinciroli
Facultad de Ciencias Exactas, Físicas y Naturales
Universidad Nacional de San Juan, Argentina
eormeno@gmail.com; fpinciroli@solus.com.ar

## Abstract

A systematic mapping protocol is a method for conducting a literature review in a rigorous and transparent way. It aims to provide an overview of the current state of research on a specific topic, identify gaps and opportunities, and guide future work. In this document, we present a systematic mapping protocol for investigating the role of the UX designer in the software development process. We define the research questions, scope, sources, search strategy, selection criteria, data extraction, and analysis methods that we will use to conduct the mapping study. Our goal is to understand how the UX designers collaborate with other stakeholders, what methods and tools they use, what challenges they face, and what outcomes they achieve in different contexts and domains.

## 1. Introduction

Our goal is to provide an overview of the current state of the art on the role of the UX designer [1] in the software development process. We also briefly introduce the concept of UX design and some of the known strategies and frameworks that support it.

User experience (UX) design is a multidisciplinary field that encompasses aspects such as usability, aesthetics, emotions, accessibility [2], and user satisfaction in relation to a product or service. UX design is increasingly recognized as a key factor for the success of software products, especially in competitive and dynamic markets. However, UX design also poses several challenges for software development teams, such as how to integrate UX activities with agile methods, how to communicate and collaborate with different stakeholders, how to measure and evaluate UX outcomes, and how to cope with changing user needs and expectations.

Several strategies and frameworks have been proposed to address these challenges and to guide software development teams in applying UX design principles and practices. Some examples are Lean UX [3], Agile UX, User-Centered Design (UCD) [4] and Design Thinking [5]. These approaches vary in terms of their underlying philosophy, methods, techniques, tools, roles, and artifacts [6]. However, there is no consensus on which approach is best suited for different contexts and situations, nor on how to effectively implement them in real-world projects.

Therefore, there is a need for a systematic mapping study that can provide a comprehensive overview of the existing literature on the role of the UX designer in the software development process, and identify the main trends, gaps, and challenges in this domain.

Evidence-Based Software Engineering (EBSE) aims to convert the need for information into an answerable question, tracking down the best evidence with which to answer that question and critically appraising the evidence for its validity. Kitchenham et al. affirm that EBSE intends "to provide the means by which current best evidence from research can be integrated with practical experience and human values in the decision making process regarding the development and maintenance of software" [7]. In this document we detail the planning phase of a Systematic Mapping Study (SMS), used to structure the findings on a research area, based on the guidelines from Petersen et al. [8].

The rest of this article is structured as follows: in section 2 we describe the research method, section 3 presents the strategy to deal with validity threats, and, finally, section 4 offers our conclusions.

## 2. Research method

### 2.1. Goal and research questions

This work aims to identify and classify UX frameworks, tools, project characteristics, the strategies for implementing the static design (in case they are used) and the skills of the team regarding the UX design.

Particularly we will consider:

- UX design approaches and frameworks.
- UX design tools.
- The various methods and practices used to incorporate UX design throughout the software development lifecycle.
- Considering our focus on contemporary full-stack applications, our review will be confined to studies conducted post-mid-2010s.
- The skills of the team regarding UX design.
- The front-end technologies in the industry.

A set of Research questions (RQ) has been designed to accomplish this goal (see Table 1). Furthermore, a set of publication questions (PQ) has been included to characterize the bibliographic and demographic space (Table 2).

Table 1. Research questions

| RQ# | Research question | Description |
| --- | --- | --- |
| RQ1 | What UX design approaches have been mentioned? | A list of UX design approaches: Lean UX, Agile UX, User-Centered Design (UCD), etc. |
| RQ2 | What UX tools have been mentioned? | A list of tools used for design: Sketch, Figma, Canva, etc. |
| RQ3 | What are the methodologies and practices employed to integrate UX design into the software development process? | This question seeks to understand the various methods and practices used to incorporate UX design throughout the software development lifecycle. |
| RQ4 | Who is primarily responsible for executing UX design tasks during software development: specialists or the software developers themselves? | This question aims to explore whether UX design tasks are typically handled by specialized UX designers or if they are carried out by the software developers themselves. |
| RQ5 | What frontend technologies are mentioned? | A list of UI frameworks: blade, React, Vue, Angular, Flutter |

Table 2. Publication questions

| PQ# | Publication question | Description |
| --- | --- | --- |
| PQ1 | Where the studies had been published? | To know the distribution of studies by type of venue: conferences, journals, or workshops. |
| PQ2 | How has the quantity of studies evolved? | Publications per year. |
| PQ3 | What are the authors' affiliations? | Classify the affiliations into two categories: academy or industry. We will consider the affiliations of all the authors. |
| PQ4 | Which are the most active countries? | Considering the author's affiliations (all authors). |

## 2.2. Search strategy and study selection

The selected search strategy includes two approaches to look for the primary studies. The first one is an automatic search on the most important online sources for scientific studies (digital libraries and databases). The second one, used in order to ensure the completeness of our set of studies, is the snowballing technique [9]. Figure 1 shows these strategies.

Figure 1. Search and selection process

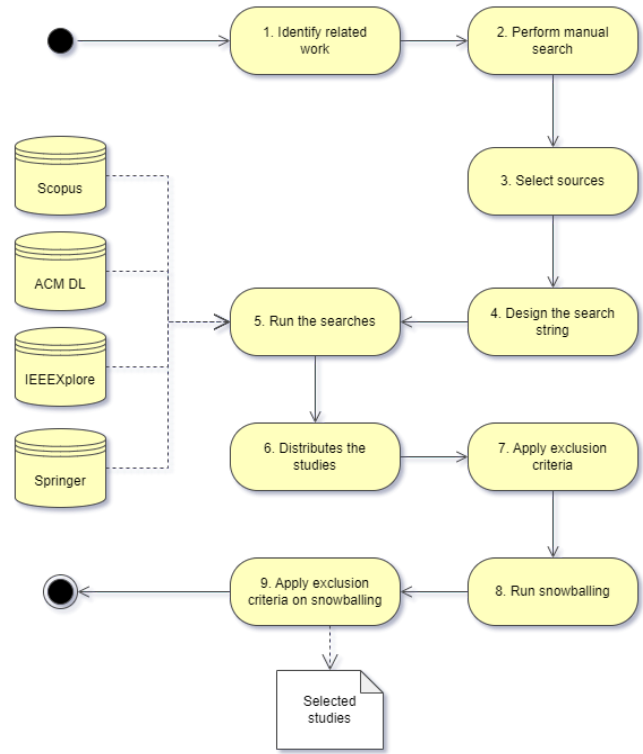

The study selection strategy will include the classification and revision of every study in the set of retrieved works, aiming to select those relevant papers, regarding the RQs, as presented in Figure 1 above.

The activities are as follows:

*Activity #1: Identify related works*

We will start our study identifying the related works previously conducted, as recommended by Petersen et al. [8], since this will also help us to adjust the focus of our study. As we are conducting a SMS, a type of secondary study, we will only consider as related work other secondary works (SMS or SLR) previously published.

*Activity #2: Perform manual search*

We have planned to conduct a manual of those related works, in order to calibrate our study.

*Activity #3: Select sources*

The electronic databases of scientific articles selected for this study are Scopus, IEEE Xplore, ACM digital library, and Springer, as they are cited repeatedly in SMS reports and guidelines [10] [11] [12] [13].

*Activity #4: Design the search string for each source*

The selection of search terms, primarily applied to titles, abstracts, and keywords, aligns with the categories delineated by the PICO method (Petersen et al. [14]):

- **Population**: The aim is to characterize User Experience (UX) or User Interface (UI) design in the context of software development. Consequently, the chosen terms encompass "UX" and "UI".
- **Intervention**: This category pertains to areas of software development, hence the term "software" has been selected.

- **Comparison**: Given the primarily descriptive nature of our study, we have not designated terms for comparison.
- **Outcome**: While specific terms for this category have not been established, we will seek evidence as indicated in the Research Questions (Table 1).

The specific search strings for each database are the following ones:

Table 3. Search strings

| Database | Search string |
|---|---|
| Scopus | `TITLE-ABS-KEY ( software AND development AND ( ( ux OR ui ) AND design ) ) AND PUBYEAR > 2014` |
| IEEE Xplore | `(("Index Terms":software) AND ("Index Terms":development) AND (("Index Terms": ux OR "Index Terms":ui) AND "Index Terms": design))` |
| ACM DL | `"query": { Title:(software AND development AND ((ux OR ui) AND design)) OR Abstract:(software AND development AND ((ux OR ui) AND design)) OR Keyword:(software AND development AND ((ux OR ui) AND design)) }`<br>`"filter": { Article Type: Research Article, E-Publication Date: (01/01/2015 TO 12/31/2024), ACM Content: DL }` |
| Springer Link | `software AND development AND ((ux OR ui) AND design)` |

*Activity #5: Run the searches*

The searches are executed, and the results collected. These results will contain duplicates that must be eliminated by applying these rules:

1. Expanded works (or expanded versions): keep the last one.
2. Duplicated works: depending on the source, following this priority order: Scopus (since it offers the most detailed information), followed by IEEE Xplore, then Springer, and, finally, ACM DL (because it does not retrieve the abstracts of the studies) [13].

*Activity #6: Distribute the studies*

The retrieved studies will be distributed among four researchers as Table 4 shows. Notice that we ensure that every single work will be examined by two different researchers, in order to reduce bias.

Table 4. Distribution of studies.

|  | Studies | | | |
|---|---|---|---|---|
| **Researcher** | **0% - 25%** | **26% - 50%** | **51% - 75%** | **76% - 100%** |
| **R1** | X | X | | |
| **R2** | | X | X | |
| **R3** | | | X | X |
| **R4** | X | | | X |

The individual selection of studies made by each researcher will be consolidated into a unique set of studies. Differences among researchers will be solved by using the following criteria [8]:

Table 5. Criteria to resolve disagreements.

|  |  | Researcher 1 | | |
| --- | --- | --- | --- | --- |
|  |  | **Include** | **Uncertain** | **Exclude** |
| **Researcher 2** | **Include** | A | B | D |
|  | **Uncertain** | B | C | E |
|  | **Exclude** | D | E | F |

- A & B: the study is included.
- E & F: the study is excluded.
- C & D: the paper is read in full and qualified again until obtaining A, B, E or F.

*Activity #7: Apply exclusion criteria*

The researchers will independently review the studies they have been assigned to and they will decide whether the studies are relevant or not, by only reading their title and abstract and then applying the exclusion criteria (EC). The set of retrieved studies will be then filtered by applying the exclusion criteria described below, in Table 6.

Table 6. Exclusion criteria.

| EC# | Description |
| --- | --- |
| EC1 | The study is not written in English. |
| EC2 | The study venue is not a conference, workshop, or journal. |
| EC3 | The study is not peer-reviewed. |
| EC4 | Short papers (four pages or less). |
| EC5 | The focus is not on UX design. |
| EC6 | The study doesn't discuss UX design as a valuable or specific topic. |
| EC7 | The study is before 2015 |

*Activity #8: Run snowballing*

Resulting articles will be considered as "seed works" to be used on a forward and a backward snowballing technique, following the guidelines proposed by Wohlin [9]. The motivation for running a forward and backward snowballing complementary search aims to complement the automatic search and to collaborate with the search strings refinement.

## 2.3. Data extraction form

Relevant data are extracted from the set of selected studies to answer the eight RQs and the four PQs. Data is stored into a spreadsheet with the format shown in Table 7 (Data Extraction Form, DEF) and in Table 8.

Table 7. Data extraction form for RQ.

| Study #ID | RQ1 | RQ2 | RQ3 | RQ4 | RQ5 |
|---|---|---|---|---|---|
| Study #1 | | | | | |
| Study #2 | | | | | |
| … | … | … | … | … | … |
| Study #n | | | | | |
| Accepted values | PM approach names | PM life cycle model names | Characteristic names | Characteristic names | Framework names |
| | (text) | (text) | (text) | (text) | (text) |

The choice of presentation format is contingent on the potential quantity of results: a bar chart is preferred for larger data sets, while a pie chart is utilized when the results are anticipated to be fewer in number.

Table 8. Data extraction form for PQ.

| Study #ID | PQ1 | PQ2 | PQ3 | PQ4 |
|---|---|---|---|---|
| Study #1 | | | | |
| Study #2 | | | | |
| … | … | … | … | … |
| Study #n | | | | |
| Accepted values | Fora names | Year of publication | Academia industry research center | Country names |
| | (text) | (text) | (text) | (text) |

## 3. Threats to validity

In order to minimize the impact of the validity threats categorized by Petersen et al. [8] that could affect our study, we present them with the corresponding mitigation actions:

*Descriptive validity*

This validity seeks to ensure that observations are objectively and accurately described.

- We have structured the information to be collected by means of a couple of Data Extraction Forms, for RQs and PQs, presented in Table 7 and Table 8, to support an uniform recording of data and to objectify the data extraction process.
- Besides, all the researchers will participate in an initial meeting, aimed at unifying concepts and criteria, answer any question and to demonstrate (by examples) how to conduct the data

- extraction process.
- We will also make our data extraction form.

*Theoretical validity*

The theoretical validity depends on the ability to get the information that it is intended to capture.

- We will start with a search string (Table 3) tailored for the four most popular digital libraries on computer sciences and software engineering online databases.
- An expert will provide a set of articles to verify if they are retrieved with the search string.
- A set of exclusion criteria (Table 6) to objectivize the selection process have been defined.
- We will distribute the studies among four researchers, working independently and, with an overlap of studies that ensures that each study is reviewed by at least two researchers (Table 4).
- We will combine two different search methods: an automatic search and a snowballing, to diminish the risk of not finding all the available evidence.
- It could have a minimal impact due to the selection of articles written in English and the discard of other languages.

*Generalizability*

This validity is concerned with the ability to generalize the results to the whole domain.

- Our set of RQs is general enough in order to identify and classify the findings on aspect-oriented software development methodologies regardless of specific cases, type of industry, etc. [14].

*Interpretive validity*

This validity is achieved when the conclusions are reasonable given the data.

- At least two researchers will validate every conclusion.
- Two researchers, experienced on the problem domain, will help us with the interpretation of data.

*Repeatability*

The research process must be detailed enough in order to ensure it can be exhaustively repeated.

- We have designed this protocol sufficiently detailed to allow us to repeat the process we have followed.
- The protocol, as well as the results of the study, will be published online, so other researchers can replicate the process and, hopefully, corroborate the results.

## 4. Conclusions

In conclusion, we have meticulously developed a systematic mapping study (SMS) protocol that aims to provide an overview of the current state of the art on the role of the UX designer in the software development process. This protocol, which is currently being implemented, is grounded in the guidelines published by Petersen et al. [8], ensuring a rigorous, transparent, and repeatable approach to our ongoing study.

Our team's strict adherence to these guidelines in the planning and execution of this SMS protocol has allowed us to mitigate potential threats to validity as much as possible. We believe that this rigorous approach will not only lend credibility to our eventual findings but also set a benchmark for future studies in this field.

As we continue to implement this protocol, we anticipate that our work will shed light on the pivotal role of UX designers in shaping the software development process. We also hope that our exploration of various strategies and frameworks supporting UX design will inspire further research and innovation in this field. Ultimately, we aim to contribute to the creation of more user-centric software solutions through our ongoing study.